\documentclass[twocolumn,
    aps,
    prb,
    superscriptaddress,
    longbibliography]{revtex4-2}
\usepackage{graphicx}
\usepackage{color}
\usepackage{amsmath}
\usepackage[export]{adjustbox}
\usepackage{enumitem}
\usepackage{amssymb}
\usepackage{hyperref}
\usepackage[normalem]{ulem}
\usepackage{adjustbox}
\usepackage{physics2}
\usephysicsmodule{ab,braket}

\hypersetup{
	colorlinks = true,
	linkcolor = blue,
	citecolor = blue,
	urlcolor  = blue,
}

\begin{document}

\title{Small Quantum Low Density Parity Check Codes for Near-Term Experiments}
\author{Christian Kraglund Andersen}
\thanks{c.k.andersen@tudelft.nl}
\affiliation{QuTech and Kavli Institute of Nanoscience, Delft University of Technology, 2628 CJ, Delft, The Netherlands}
\author{Eliška Greplová}
\thanks{e.greplova@tudelft.nl}
\affiliation{QuTech and Kavli Institute of Nanoscience, Delft University of Technology, 2628 CJ, Delft, The Netherlands}

\begin{abstract}
    It is widely accepted that quantum error correction is essential for realizing large-scale fault-tolerant quantum computing. Recent experiments have demonstrated error correction codes operating below threshold, primarily using local planar codes such as the surface code and color code. In parallel, theoretical advances in quantum low-density parity-check (LDPC) codes promise significantly lower overheads, albeit at the cost of requiring non-local parity checks. While these results are encouraging, implementing such codes remains challenging for near-term experiments, creating obstacles to holistic benchmarking of hardware architectures capable of supporting long-range couplers. In this work, we present a simple construction recipe for small quantum LDPC codes based on recent developments in the field. Our codes are approximately twice as efficient as comparable surface codes, yet require only weight-four parity checks, which simplifies experimental realization compared to other quantum LDPC codes. We provide concrete proposals for implementations with superconducting qubits in flip-chip architectures and with semiconductor spin qubits using shuttling-based approaches.
\end{abstract}

\maketitle

\section{Introduction}

Quantum error correction (QEC) is a key ingredient for fault-tolerant quantum computing. Recently, there has been an immense amount of progress in experimental verification of the suppression of quantum errors using QEC~\cite{ryananderson2021realtime,abobeih2021faulttolerant,krinner2022repeated,sundaresan2023demonstrating,zhao2022surface,acharya2022suppressing,bluvstein2023logical,google2025_belowthreshold,postler2024steane,putterman2025hardware,sivak2023realtime,lacroix2025colorcode}. In particular, the surface code has drawn a lot of attention and in Ref.~\cite{google2025_belowthreshold} the operation a surface code well below the error correction threshold was demonstrated. There are a number of reasons why the surface code is often considered the de-facto standard for QEC. From a practical point of view, the community knows how to built quantum devices that support the surface code~\cite{fowler2012surface, versluis2017scalable, andersen2020repeated, acharya2022suppressing, marques2022logical, google2025_belowthreshold}. Similarly, there are well understood architectures for building large-scale fault tolerant quantum computers using the surface code~\cite{fowler2012surface, litinski2019game, lee2021even, gidney2025factor} which readily allow for the estimation of physical resources for practical quantum algorithms. From a theoretical point of view, the surface code also has the optimal overhead allowed for any local code in two dimensions~\cite{bravyi2010tradeoffs}.

On the other hand, there has recently been a growing interest in quantum low-density parity-check (LDPC) codes which may have favorable scaling compared to the surface code and other local codes. It was mathematically shown that quantum LDPC codes that are asymptotically \textit{good} exist~\cite{hastings2021fiber, panteleev2021almostlinear, panteleev2021degenerate, breuckmann2021balanced, leverrier2022quantumtanner, dinur2022goodqldpc, panteleev2022asymptotically, lin2022losslessexpander}. Here, a \textit{good} code is a quantum code where the rate, meaning the number of logical qubits $k$ per physical qubits $n$, is always finite as a function of the code distance $d$ as $d \rightarrow \infty$. However, real quantum devices do not operate in the asymptotic limit. Thus, the relevant question for near-term experiments is whether we can construct quantum LDPC codes with a smaller overhead than the surface code for small and intermediate scale quantum systems. An important step towards practical quantum LDPC codes was taken in Ref.~\cite{bravyi2024highthreshold}, where a quantum code, known as a bivariate bicycle (BB) code, with the properties of $[[n{=}144,k{=}12,d{=}12]]$ was presented. We will refer to this code as the BB6-144, where the BB6 part signifies to the fact that the code features weight-6 stabilizers as opposed to the surface code which relies on weight-4 stabilizers. Similar codes with different requirements and constraints were presented in Refs.~\cite{voss2024multivariate, shaw2025morphing, ye2025ionchain, yang2025planar}. Additionally, in Ref.~\cite{yoder2025tourdegross} it was shown that if the BB6-144 code is used in a fault tolerant architecture, it could reduce the total resources for practical algorithms by roughly one order of magnitude compared to a surface code architecture. On the other hand, there is still a number of practical challenges to realize a BB6-144 code with a real quantum device. Firstly, the code requires 288 physical qubits when including auxiliary qubits which is currently beyond most high-performing quantum devices that also support fast mid-circuit measurements. Secondly, any quantum LDPC code will naturally require long range interactions and it still remains an open question for many experimental platforms how to best implement these interactions. Another important aspect of any QEC code is that of decoding. The surface code can be efficiently decoded by the minimal-weight perfect-matching  algorithm~\cite{higgott2022pymatching}. In contrast, a general quantum LDPC code requires a more involved decoding algorithm based, for example, on a belief propagation algorithm. Finally, weight-6 stabilizers of the BB6 codes imply that the connectivity for each qubit must be at least 6 which adds extra constraint on the device design although modifications can be employed to reduce the required connectivity~\cite{shaw2025morphing}. A first step towards real world experiments was presented in Ref.~\cite{wang2025lowoverhead}, where stabilizers were removed by hand such that the code becomes experimentally feasible. The error rate in this experiment was however not below the threshold for fault tolerance and the experiment did not yet demonstrate a clear advantage over surface codes.

In this work, we present small quantum LDPC codes based on Ref.~\cite{voss2024multivariate} that may be implemented more readily in near-term experiments. The goal is to enable experiments to test the components needed for large scale quantum LDPC codes such as long range interactions and thereby accelerate the time-scale until quantum LDPC codes will become practical for large scale quantum computing. To keep the implementations as simple as possible, we will focus on codes that rely only on weight-4 parity checks. While this choice simplifies the topology of the device, it also enables efficient error extraction circuits and efficient decoding using the MWPM decoder. The codes presented here offer an overhead-reduction of roughly a factor of 2 compared to the surface code and we observe a favorable scaling of the code rate when compared to the surface code. We also discuss how to implement logical gates in this circuit. Finally, we discuss how these codes may be implemented in solid-state platforms for near-term experiments. Specifically, we discuss how a simple flip-chip architecture~\cite{rosenberg2017_3d, kosen2022building, norris2025_multimodule} may enable the implementation for superconducting qubits while, for spin qubits, we discuss an implementation based on long-range shuttling~\cite{fujita2017coherentshuttle, mills2019shuttlingcharge, yoneda2021coherentspintransport, zwerver2023shuttlespin, vanriggelen-doelman2024germanshuttle, desmet2025highfidelity, katiraeefar2025evolutionary}. 

\section{Code construction}
\label{sec:code}

We begin by reviewing the framework of Ref.~\cite{voss2024multivariate} which constructs general bicycle codes. The key objects to construct these codes are the matrices $A$ and $B$ defined as a sum of $W_A$ and $W_B$ terms, respectively, where $W_A$ and $W_B$ are small integers, such that
\begin{align}
    A = \sum_{i=1}^{W_A} A_i, && B = \sum_{i=1}^{W_B} B_i. \label{eq:defAB}
\end{align}
The terms for both $A_i$ and $B_i$ are defined to be a power a single matrix, i.e., $A_i = {a_i}^{p_i}$ and $B_i = {b_i}^{q_i}$ with $p_i$ and $q_i$ integers. Now, the key step is to define the matrices $a_i$ and $b_i$. First we define two integers $l$ and $m$ which will later set the size of our code. For each of these integers, we define the shift matrices $S_l$ and $S_m$. The shift matrix of size $n$ is simply a matrix where the $i$th row has a single unit entry at the $(i+1)\text{mod}\,n$-th column. For example, we get
\begin{align}
    S_3 = \begin{pmatrix}
        0 & 1 & 0 \\
        0 & 0 & 1 \\
        1 & 0 & 0
    \end{pmatrix}.
\end{align}
We can now define three matrices 
\begin{align}
    x = S_l \otimes I_m, && y = I_l \otimes S_m, && z = S_l \otimes S_m,
\end{align}
where we choose $a_i$ and $b_i$ to be either $x$, $y$ or $z$. For example, we can set
\begin{align}
    A &= x + y^2, \label{eq:AB_dist3_A} \\
    B &= x^2 + z^4, \label{eq:AB_dist3_B}
\end{align}
for $W_A = W_B = 2$. Keep in mind that we are here formally working with the binary field $\mathbb{F}_2 = \lbrace 0,1 \rbrace$, which means that the summation in Eqs.~\eqref{eq:defAB}, \eqref{eq:AB_dist3_A} and~\eqref{eq:AB_dist3_B} must be done modulo 2.

\begin{figure}[t]
    \includegraphics[width=\linewidth]{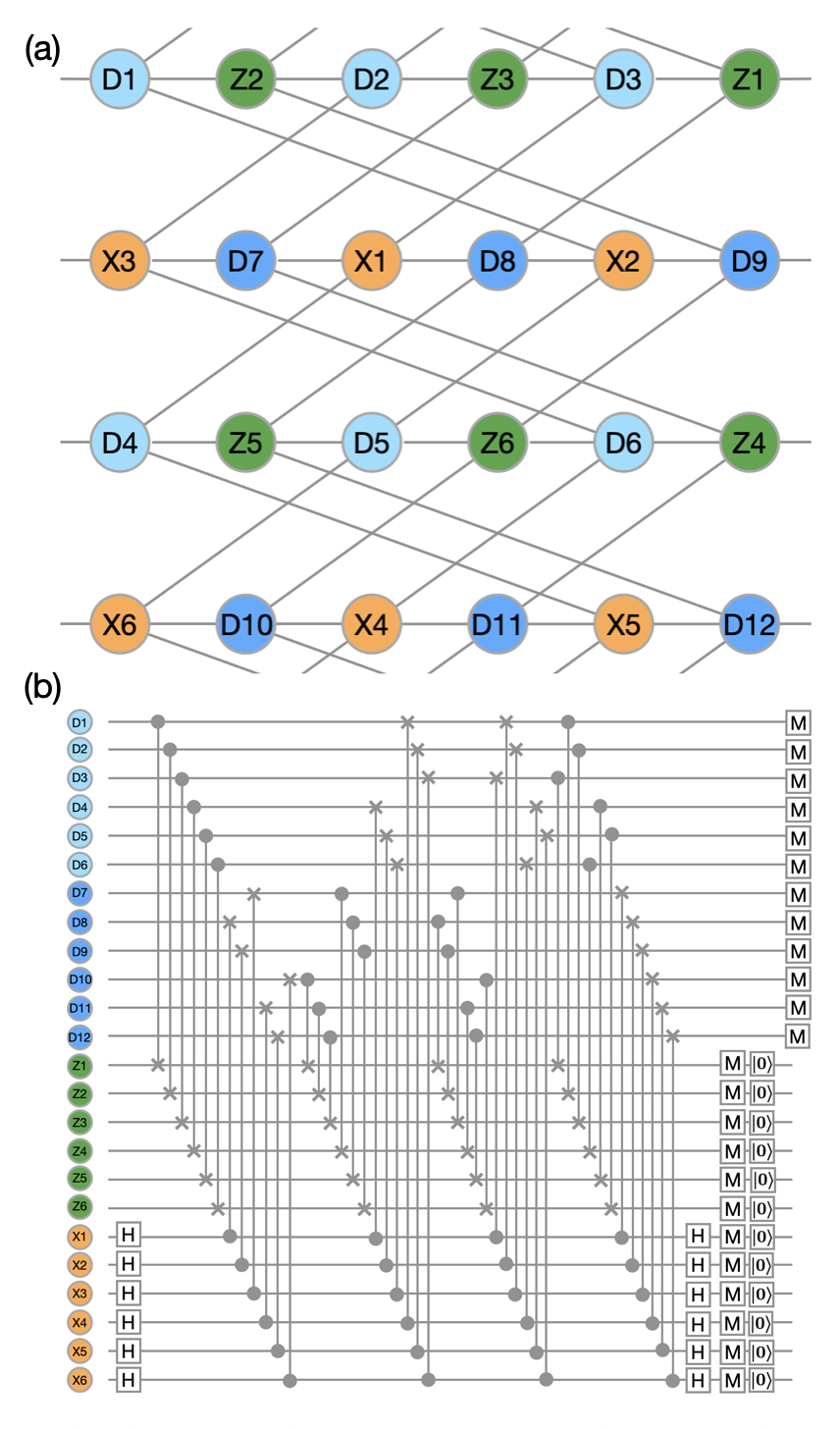}
    \caption{(a) The visualization of the [[12,2,3]] code. Data qubits are in blue, $Z$ parity checks in green and $X$ parity checks in orange. The grey lines represent the code connectivity. The open lines symbolize periodic boundary conditions. (b) The circuit representation of the one cycle of the code.}
    \label{fig:code_overview}
\end{figure}

With the matrices $A$ and $B$ defined, we can construct the the parity checks matrices by horizontally stacking $A$ and $B$ as well as their transpose. Specifically we have
\begin{align}
    H_X &= [A \vert B], \\
    H_Z &= [B^T \vert A^T],
\end{align}
where the rows in $H_X$ and $H_Z$ correspond to the $X$-stabilizers and $Z$-stabilizers, respectively. We observe that the number of columns in $H_X$ and $H_Z$ sets the number of physical data qubits $n = 2lm$. Note that the number of stabilizers are also equal to $2lm$, so it may be surprising that the code supports potentially multiple logical qubits. The trick here is that the stabilizers in $H_X$ and $H_Z$ are not linearly independent, which may appear inefficient but the redundancy allows the code to maintain the low-density property of a LDPC code. As pointed out in Ref.~\cite{bravyi2024highthreshold}, we can therefore find the number of logical qubits $k$ by
\begin{align}
    k = 2\,\text{dim}[\text{ns}(A) \cap \text{ns}(B) ], \label{eq:k}
\end{align}
where \textit{dim} refers to the dimension and \textit{ns} defines the null-space of the matrix. Similarly, we can also find the code distance as
\begin{align}
    d = \text{min}\lbrace |v|\,\text{for}\, v \in \text{ns}(H_X) \backslash \text{rs}(H_Z)\rbrace,
\end{align}
where \textit{rs} refers to the row-space and $\vert \cdot \vert$ is the Hamming weight. The code distance can be found through an efficient numerical optimization~\cite{shaw2025morphing}. An interesting observation is that the weight of the stabilizers in both $H_X$ and $H_Z$ are by construction equal to the integer number $W=W_A + W_B$, meaning the more terms we include in $A$ and $B$, the higher weight stabilizers we have.

Let us now consider an explicit example of the quantum LDPC code we described above. We consider a code with $l=2$ and $m=3$ with $A$ and $B$ defined as in Eqs.~\eqref{eq:AB_dist3_A} and~\eqref{eq:AB_dist3_B}. In this case, we find a code with the properties of $[[n,k,d]] = [[12,2,3]]$ and with the check matrices given by
\begin{align}
H_Z = \left[
\begin{array}{cccccccccccc}
1 & 0 & 1 & 0 & 0 & 0 & 0 & 1 & 0 & 1 & 0 & 0 \\
1 & 1 & 0 & 0 & 0 & 0 & 0 & 0 & 1 & 0 & 1 & 0 \\
0 & 1 & 1 & 0 & 0 & 0 & 1 & 0 & 0 & 0 & 0 & 1 \\
0 & 0 & 0 & 1 & 0 & 1 & 1 & 0 & 0 & 0 & 1 & 0 \\
0 & 0 & 0 & 1 & 1 & 0 & 0 & 1 & 0 & 0 & 0 & 1 \\
0 & 0 & 0 & 0 & 1 & 1 & 0 & 0 & 1 & 1 & 0 & 0
\end{array}
\right]
\end{align}
and
\begin{align}
    H_X = \left[
\begin{array}{cccccccccccc}
        0 & 0 & 1 & 1 & 0 & 0 & 1 & 1 & 0 & 0 & 0 & 0 \\
        1 & 0 & 0 & 0 & 1 & 0 & 0 & 1 & 1 & 0 & 0 & 0 \\
        0 & 1 & 0 & 0 & 0 & 1 & 1 & 0 & 1 & 0 & 0 & 0 \\
        1 & 0 & 0 & 0 & 0 & 1 & 0 & 0 & 0 & 1 & 1 & 0 \\
        0 & 1 & 0 & 1 & 0 & 0 & 0 & 0 & 0 & 0 & 1 & 1 \\
        0 & 0 & 1 & 0 & 1 & 0 & 0 & 0 & 0 & 1 & 0 & 1 
    \end{array}
\right].
\end{align}
These matrices correspond to the $Z$ stabilizers given by
\begin{align}
    S_{Z1} &= Z_1 Z_3 Z_8 Z_{10}, \label{eq:S_Z1}\\
    S_{Z2} &= Z_1 Z_2 Z_9 Z_{11},\\
    S_{Z2} &= Z_2 Z_3 Z_7 Z_{12},\\
    S_{Z4} &= Z_4 Z_6 Z_7 Z_{11},\\
    S_{Z5} &= Z_4 Z_5 Z_8 Z_{12},\\
    S_{Z6} &= Z_5 Z_6 Z_9 Z_{10}
\end{align}
and $X$ stabilizers given by
\begin{align}
    S_{X1} &= X_3 X_4 X_7 X_8, \\
    S_{X2} &= X_1 X_5 X_8 X_9, \\
    S_{X3} &= X_2 X_6 X_7 X_9, \\
    S_{X4} &= X_1 X_6 X_{10} X_{11}, \\
    S_{X5} &= X_2 X_4 X_{11} X_{12}, \\ 
    S_{X6} &= X_3 X_5 X_{10} X_{12}. \label{eq:S_X6}
\end{align}
Here $Z_i$ and $X_i$ refers to the Pauli $Z$ and $X$ operators for data qubit $i$, respectively. Notice that there is a natural separation of the qubits into two halfs based on the construction of $H_X$ and $H_Z$. We refer to the first $n/2$ qubits as the \textit{left} data qubits and the second half as the \textit{right} data qubits. Each stabilizer connects, by construction, to exactly 2 left data qubits and 2 right data qubits. While this statement is true independent of the choice of $A$ and $B$ for fixed $W_A$ and $W_B$, we have picked here a code with the additional property that the left data qubits separate into two sets for the $Z$ stabilizers and the right data qubits separate into two sets for the $X$ stabilizers. This choice allows us to make a simple geometric toric layout of the code, see Fig.~\ref{fig:code_overview}(a). In general for larger codes, this geometric construction is possible as long as we can separate each half into $l$ disconnected sets. At this point, we already see a modest improvement over surface codes as we only require 12 physical data qubits for 2 logical qubits. The original surface code of distance $d$ requires $d^2+(d-1)^2$ physical qubits per logical qubit meaning 25 data qubits are needed for a distance-3 code~\cite{dennis2002topological, fowler2012surface}. On the other hand, the rotated version of the surface code requires only $d^2$ data qubits~\cite{horsman2012surfacecodelattice, tomita2014lowdistance}.  Thus, when including the auxiliary qubits, the $[[12,2,3]]$ code presented here requires 24 physical qubits when including the auxiliary qubits while the rotated surface code requires 34 physical qubits for 2 distance-3 logical qubits. This code is, incidentally, analogous to the twisted toric code of Ref.~\cite{breuckmann2021balanced}. As we will discuss in Sec.~\ref{sec:scaling}, the reduced overhead compared to the surface code is not simply a constant scaling factor as for other local codes such as the color code~\cite{bombin2007branyons, bombin2007optimalresources, gidney2023colorcodetools, lacroix2025colorcode}. Rather, the code construction that we have reviewed here provides an improved scaling compared to local codes and via the code repository in Ref.~\cite{GitHub} we provide the tool for anyone to construct a custom code using the code construction from Ref.~\cite{voss2024multivariate}. 

A final important aspect of the code construction is the identification of the logical operators for the encoded qubits. From Eq.~\eqref{eq:k}, we readily know how many logical qubits we have, but we must find a pair of operators $Z_{Li}$ and $X_{Li}$ for each $i \in [1, \ldots , k]$ such that we have the anti-commutators
\begin{align}
    \lbrace Z_{Li}, X_{Li} \rbrace = 0
\end{align}
as well as the commutators
\begin{align}
    [Z_{Li}, X_{Lj}] = 0 && \text{for } i \neq j.
\end{align}
Additionally, the logical operators must commute with the stabilizers
\begin{align}
    [Z_{Li}, S_{Xj}] = 0, && [X_{Li}, S_{Xj}] = 0,
\end{align}
for all $j$. To identify such pairs of logical operators, we begin by considering the \textit{centralizer} of the stabilizer group $S$ generated by all the stabilizers $S_{Xi}$ and $S_{Zi}$~\cite{nielsen2000quantum, wilde2009logicaloperators}. This centralizer, denoted $\mathcal{C}(S)$, consists of all Pauli operators that commute with every element of the stabilizer group $S$. By construction, any operator in $\mathcal{C}(S)$ preserves the codespace. However, operators in $S$ act trivially on the logical subspace, so we are interested in the so-called quotient space $\mathcal{C}(S) \text{ mod } S$, which captures the nontrivial logical action. This quotient space should have the dimension $2k$, corresponding to the $2k$ logical operators (one pair for each logical qubit). A basis for this space can be obtained by first computing a generating set for $\mathcal{C}(S)$, and then removing any elements that lie within the stabilizer group itself. To construct explicit logical operators, we select $2k$ elements from this space and organize them into $k$ symplectic pairs ${ (X_{L1}, Z_{L1}), \ldots, (X_{Lk}, Z_{Lk}) }$ such that the commutators and anti-commutators above hold. This pairing can be achieved via a symplectic Gram-Schmidt orthogonalization process as detailed in Ref.~\cite{wilde2009logicaloperators}. The resulting operators form a complete set of logical $X_{Li}$ and $Z_{Li}$ operators for $i = 1, \ldots, k$, which act faithfully and independently on the $k$ encoded qubits while commuting with the stabilizer group. For the $[[12,2,3]]$ code with the stabilizers defined in Eqs.~\eqref{eq:S_Z1}-\eqref{eq:S_X6}, we arrive at the logical operators
\begin{align}
    Z_{L1} &= Z_1 Z_2 Z_3 Z_4 Z_5 Z_6, && Z_{L2} = Z_1 Z_3 Z_5 Z_6 Z_7, \label{eq:logical_z} \\
    X_{L1} &= X_1 X_2 X_3, && X_{L2} = X_1 X_2 X_4 X_5 X_7 X_{10}. \label{eq:logical_x}
\end{align}
As always, these logical operators are not unique as any multiplication of a stabilizers with the logical operator will yield an equivalent logical operator. For example, it is straight forward to notice that $X_{L1} = X_4 X_5 X_6$ is also a valid $X$-operator for the first logical qubit. Since the code that we construct here are of the Calderbank-Shor-Steane-type code, we can apply the logical $X$ and $Z$ operators transversally on the code. In Sec.~\ref{sec:logicalgates}, we will return to this point to explore how other logical gates can be applied within the codespace.

\section{Performance and decoding}

\begin{figure}[t]
    \centering
    \includegraphics[width=\linewidth]{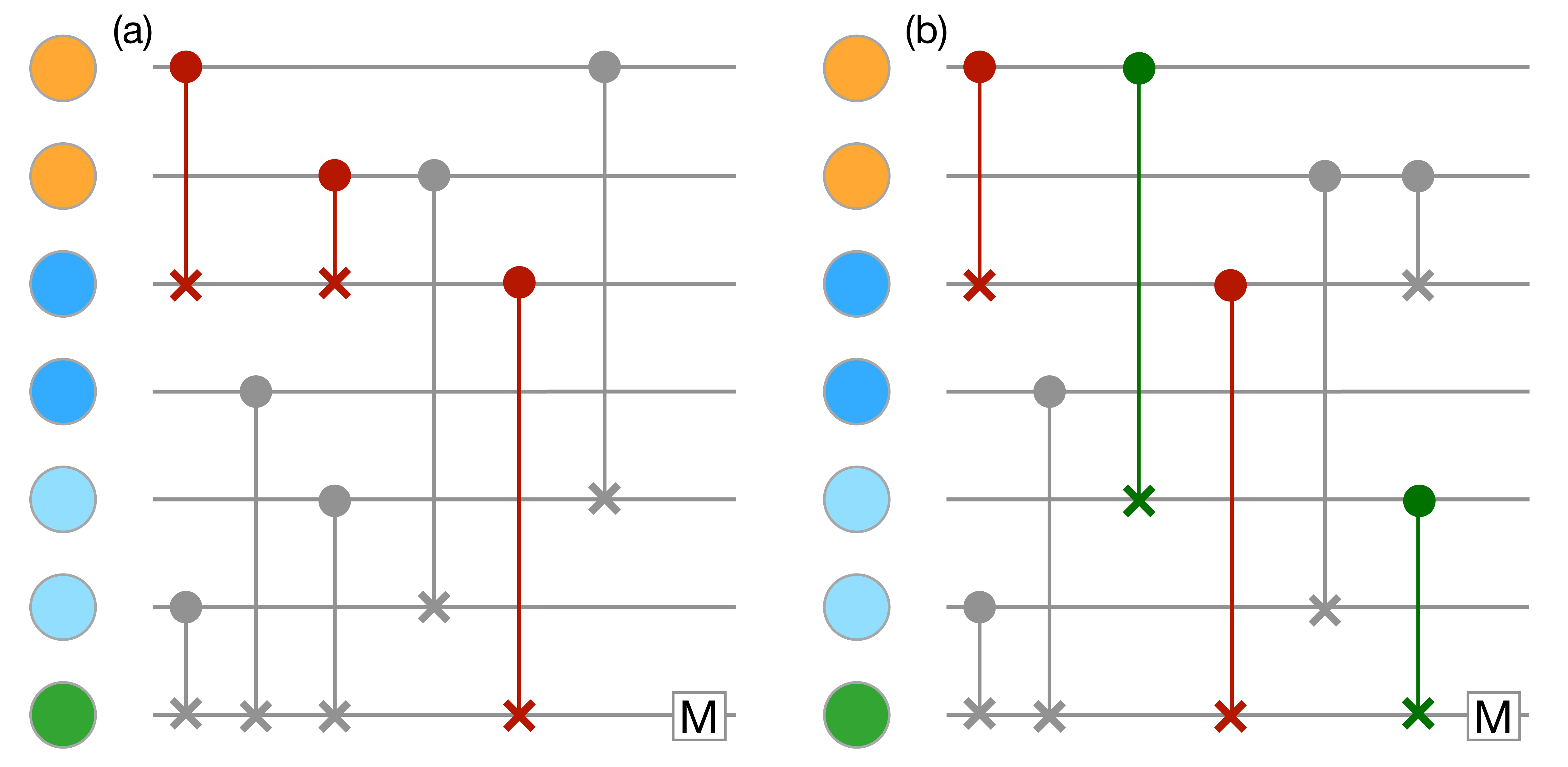}
    \caption{(a) If the error extraction circuit is not constructed correctly, the circuit will create an entangled state between the $Z$ and $X$ stabilizer auxiliary qubits mediated by the red gates. (b) In a good circuit, the entangling gates commute fully and disentangle the auxiliary qubits from each other through the green gates. Thus, the effictive stabilizer measurements fully commute.}
    \label{fig:error_circuit}
\end{figure}

\subsection{Syndrome extraction circuit}
A key challenge for general quantum LDPC codes is the design of syndrome extraction circuits that are fault-tolerant and avoid introducing correlated errors~\cite{breuckmann2021quantumldpc}. Owing to the simple weight-4 stabilizer constructions considered in this work, it is possible to explicitly specify a complete syndrome extraction circuit, as illustrated in Fig.\ref{fig:code_overview}(b), which will be of the same depth as a surface code. To emphasize the subtlety of constructing efficient syndrome extraction procedures, we show in Fig.\ref{fig:error_circuit} a concrete example that highlights the issue. Specifically, we consider a single $Z$-type auxiliary qubit and its four neighboring data qubits, along with two $X$-type auxiliary qubits coupled to the same data qubits. If one selects an arbitrary ordering of the entangling gates, such as the sequence shown in panel (a), the circuit may inadvertently create an entangled state involving the auxiliary qubits. In this situation, the measurement outcomes of the $X$- and $Z$-type stabilizers are no longer statistically independent, which can compromise fault tolerance.

This effect can be understood more precisely in two complementary ways. From a connectivity perspective, as depicted in Fig.~\ref{fig:code_overview}(a), each auxiliary qubit is coupled to exactly two left data qubits and two right data qubits. Concretely, if the $Z$-type auxiliary qubit first interacts with the two left-type data qubits before any entangling operations between those data qubits and the $X$-type auxiliary qubits have occurred, then subsequent interactions with the right-type data qubits propagate entanglement across the auxiliary qubits. Equivalently, this can be formulated in terms of the commutation relations between the controlled gates implementing the stabilizer measurements. Ideally, the sequence of entangling operations should correspond to a set of controlled-Pauli gates whose collective action commutes pairwise between $X$- and $Z$-type stabilizers. In particular, if the gates corresponding to the $X$- and $Z$-type stabilizers fail to commute, the result is an effective entangling operation between auxiliary qubits. This entanglement is manifested in the measurement statistics as nontrivial commutators between the extracted stabilizer observables, thereby violating the requirement that the syndrome bits are independent random variables conditioned only on the data qubit errors. Note that the issue discussed here is unrelated to hook error~\cite{tomita2014lowdistance} which will depend on how the logical operators are defined as well. In this work, we verify that our error extraction circuits are fault tolerant by validating the performance of the decoded error rate.

\begin{figure}[t]
    \centering
    \includegraphics[width=\linewidth]{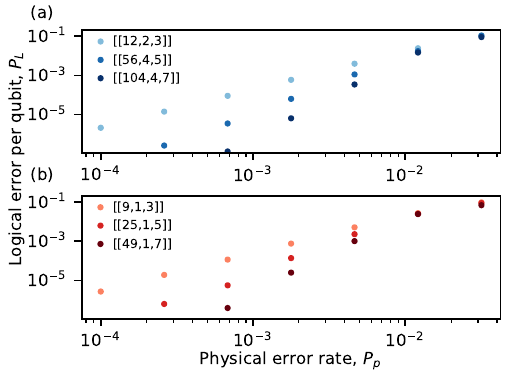}
    \caption{Logical error rate per logical qubit, $P_L$, when decoded with a minimal-weight perfect-matching decoder as a function of the physical error rate $P_p$. (a) The quantum LDPC codes constructed in the main text. (b) Rotated surface code encoding a single logical qubit.}
    \label{fig:performance_codes}
\end{figure}

\begin{figure*}[t]
    \centering
    \includegraphics[width=\linewidth]{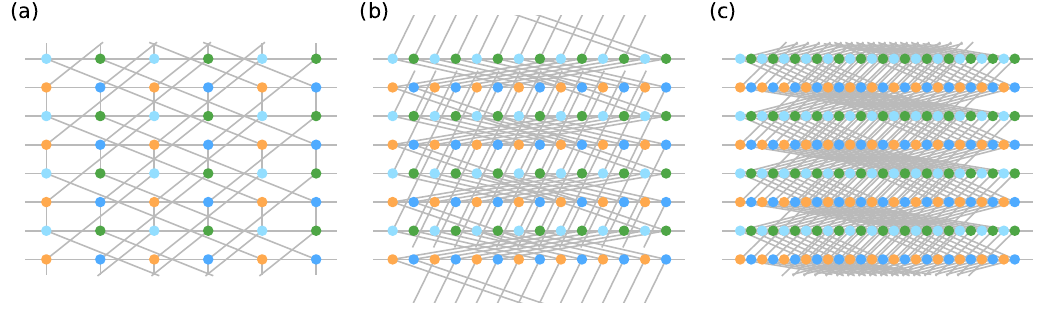}
    \caption{Larger codes illustrated with the data qubits in blue, the $X$-type auxillarity qubits in orange and the $Z$-type auxiliary qubits in green. (a) A $[[24,4,3]]$ code constructed with $l=4$ and $m=3$. (b) A $[[56,4,5]]$ code constructed with $l=4$ and $m=7$. (c) A $[[104,4,7]]$ code constructed with $l=4$ and $m=13$.}
    \label{fig:codes}
\end{figure*}

In contrast, Fig.~\ref{fig:error_circuit}(b) shows an ordering of the entangling gates that avoids the problem described above. In this circuit, the $Z$-type auxiliary qubit interacts with its left-type data qubits both at the beginning and at the end of the gate sequence. In terms of operator ordering, this arrangement preserves the commutation of the effective stabilizer measurement operators, because each data qubit interaction is bracketed in a way that prevents the controlled gates from generating unwanted commutators when conjugated through the rest of the circuit. It is straightforward to find such compatible gate orderings for the code presented here because the weight-4 stabilizers decompose naturally into two left-type data qubits and two right-type data qubits. This partitioning provides a convenient structure that ensures that the auxiliary qubits remain disentangled at the end of the circuit. This simplicity is a key advantage of the specific code construction used here.

\subsection{Decoding and error rates}

As for the error extraction circuits, the simple weight-4 stabilizers also allow us to separate the $X$-type errors and the $Z$-type errors into two separate error graphs to decode similar to a surface code. As such, we can employ the minimal-weight perfect-matching algorithm (MWPM)~\cite{fowler2012practicalprocessing, fowler2012thresholdprl}. It is known that MWPM only approximates the error model since it essentially only considers X and Z errors~\cite{higgott2023beliefmatching}. However, with usage of several error correction rounds our code does not appear to be meaningfully hindered by this limitation. We use the implementation in Ref.~\cite{higgott2022pymatching} which takes the full error correction circuit implemented in \textit{stim}~\cite{gidney2021stim} as input to construct the weights of the decoding problem. We implement the circuit as in Fig.~\ref{fig:code_overview}(b) and we include a initialization error, measurement error and a depolarization error after all gates each with an error probability $P_p$ which we refer to as the physical error. A depolarization error on single qubit corresponds to the simultaneous application of $X$, $Y$ and $Z$ errors with equal strength. Following two-qubit gates, we depolarize the full two-qubit subspace. Using \textit{stim}, we simulate the code for $d$ rounds of syndrome extraction before a final measurement of the data qubits and we decode the error syndromes to find the number of total errors $E$. Then, we can calculate the total logical error rate as $P_k = 1-(1-E/N)^{1/r}$, where $N$ is the total number of repetitions and $r$ is the number of rounds. Finally, to readily compare codes with different number of qubits $k$, we extract the logical error rate per qubit $P_L = 1-(1-P_k)^{1/k}$, see Fig.~\ref{fig:performance_codes}(a) where we use $r=d$ for all simulations. In particular, we notice that at a physical error rate $0.2\times 10^{-3}$, the logical error per qubit is below $10^{-5}$. For comparison we also extract the logical error rate for a rotated surface code with the same error model and decoder, see Fig.~\ref{fig:performance_codes}(b). We notice a very similar performance as the $[[12,2,3]]$-code. In other words, the reduced overhead of the bicycle code does not lead to a decrease in the code performance.

\subsection{Larger codes}

To demonstrate the performance of our code beyond a simple distance-3 code, we use the same construction as in Sec.~\ref{sec:code} but increase $l$ and $m$. Specifically, we pick $l=4$ and $m=3$, $7$ and $11$ to construct codes with $k=4$ and distance $d=3$, $5$ and $7$, respectively. These codes each have their own characteristic choice of $A$ and $B$. For the distance 3 code with $k=4$ in Fig.~\ref{fig:codes}(a), we have
\begin{align}
    A_{[[24,4,3]]} &= x + z^7, \\
    B_{[[24,4,3]]} &= I + y,
\end{align}
where $I$ is the identity matrix. We find that the performance per logical qubit matches fully with the $[[12,2,3]]$-code since this code is effectively a copy of two $[[12,2,3]]$-codes.

For the distance 5 code in Fig.~\ref{fig:codes}(b), we use
\begin{align}
    A_{[[56,4,5]]} &= y^6 + z^{22},\\
    B_{[[56,4,5]]} &= y + y^2.
\end{align}
In Fig.~\ref{fig:performance_codes}(a), we see that the logical error rate per logical qubit slightly outperforms the distance-5 rotated surface code in Fig.~\ref{fig:performance_codes}(b) despite the reduced overhead by a factor of $\approx 1.7$. 

Finally, we present a $k=4$ code with a code distance of $d=7$ in Fig.~\ref{fig:codes}(c) where we use
\begin{align}
    A_{[[104,4,7]]} &= x + z^{35},\\
    B_{[[104,4,7]]} &= y^4 + y^5.
\end{align}
As for the distance-3 codes, we see in Fig.~\ref{fig:performance_codes} that this distance-7 code performs similar to a distance-7 rotated error correction code. All three codes exemplified here have been found with a random search of possible codes using the code provided in Ref.~\cite{GitHub}.

\begin{figure}[t]
    \centering
    \includegraphics[width=\linewidth]{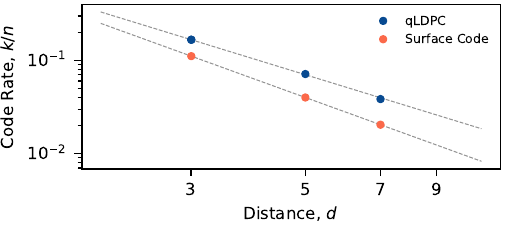}
    \caption{Comparison of the code rates for our codes and the surface code. The rate $R=k/n$ is shown as a function of the code distance $d$ for the quantum LDPC codes in Fig.~\ref{fig:codes} (blue) and for the rotated surface code (orange).}
    \label{fig:code_rate}
\end{figure}

\subsection{Scaling}
\label{sec:scaling}

The surface code is known to have the optimal scaling for any local planar error correction code as the distance increases. To be more precise, it was shown in Ref.~\cite{bravyi2010tradeoffs} that any stabilizer code defined by geometrically local interactions in two dimensions must satisfy the tradeoff $k d^2 \leq c n$,
where $k$ is the number of logical qubits, $d$ is the code distance, $n$ is the total number of physical qubits, and $c$ is a constant that depends on the locality constraints. This immediately implies that for any fixed $k$, the minimum overhead scales as
\begin{align}
    n = \mathcal{O}(d^2),
\end{align}
which is achieved asymptotically by the surface code and by the color code as well as opther local codes. Thus, if we define the rate $R = \frac{k}{n}$ we must have $R \propto d^{-2}$. As the distance $d$ increases to suppress logical error rates, the physical footprint of the code must grow quadratically in $d$ when restricted to strictly local planar couplings. Interestingly, this scaling makes quantum algorithms that provide a quadratic speed-up substantially less appealing for achieving practical quantum speed-ups~\cite{babbush2021focusbeyondquadratic, stoudenmire2024opening}.

As shown in Fig.~\ref{fig:code_rate}, the biclycle codes constructed here demonstrate a larger rate compared to the rotated surface code at the same code distance. More importantly, the scaling of $R$ with the distance $d$ is improved over the surface code. To quantify this, we performed a least-squares fit of the rate scaling with a power law of the form $R(d) = \alpha d^{-\beta}$
where $\alpha$ and $\beta$ are fit parameters. For the rotated surface code, $\beta \approx 2.0$ as expected, while for the quantum LDPC codes we observe $\beta \approx 1.7$, showing a clear advantage from the additional non-local connections enabled by long-range couplers. This effectively allows us to encode more logical qubits per unit area without sacrificing distance to the same degree as a purely local construction. Note, however, that the simple weight-4 connectivity used here means that all the constructed codes can be understood as surface code with \textit{twists} and rotations. Thus, the reduced overhead comes directly from the from non-local connections required to form the appropriate modifications of the surface code~\cite{wang2022distance_bounds_generalized_bicycle}.

It should be emphasized, however, that this result should not be over-interpreted, since there is no generic, systematic way of scaling a generalized bicycle code to arbitrarily large distance $d$. In particular, for weight-4 codes, any such generic scaling is bound to scale similar to the surface code~\cite{wang2022distance_bounds_generalized_bicycle}. For each target distance, the underlying matrices $A$ and $B$ in the code construction must be chosen by hand to satisfy the required distance and weight constraints. In particular, while in principle the quantum LDPC codes can yield large codes with finite rate and distance scaling sublinearly with $n$, practical realizations are limited by the feasibility of constructing and embedding the constituent codes in hardware, see also Sec.~\ref{sec:implementation}. Nonetheless, the observed scaling exponent $\beta < 2$ demonstrates the potential of architectures incorporating non-local couplers to overcome some of the rate limitations inherent to planar locality.

\begin{table}[t]
    \centering
    \begin{tabular}{c|l}
         Logical gate & Physical gate sequence  \\
         \hline
         $H_{L1}$ 
              & (i) $X_1 X_2 X_3 $ \\
              & (ii) $C_1 C_2 C_3$ \\
              & (iii) $CZ_{1,2}$ \\
              & (iv) $CZ_{1,3}$ \\
              & (v) $CZ_{1,4}$ \\
              & (vi) $CZ_{1,5}$ \\
              & (vii) $CZ_{1,6}$ \\
              & (viii) $CZ_{2,3}$ \\
              & (ix) $CZ_{2,4}$ \\
              & (x) $CZ_{2,5}$ \\
              & (xi) $CZ_{2,6}$ \\
              & (xii) $CZ_{3,4}$ \\
              & (xiii) $CZ_{3,5}$ \\
              & (xiv) $CZ_{3,6}$ \\
              & (xv) $\sqrt{X_1}^\dagger \sqrt{X_2}^\dagger \sqrt{X_3}^\dagger $ \\
              & (xvi) $S_4 \; S_5 \; S_6$ \\
         \hline
         $S_{L2}$ & (i) $Z_1 Z_2 Z_4 Z_6 Z_7 Z_9$ \\
          & (ii) $CZ_{7,8}$ \\
          & (iii) $CZ_{7,9}$ \\
          & (iv) $CZ_{8,9}$ \\
          & (v) $S_7 S_8 S_9$ 
    \end{tabular}
    \caption{Examples of logical Clifford gates for the $[[12,2,3]]$ code. Here, $X_i$ and $Z_i$ are the Pauli operators for qubit $i$, $S_i$ is the $S$-gate corresponding to a $\pi/2$-rotation around the $Z$-axis and $C_i$ is the Clifford gate on qubit $i$ corresponding to first a $\pi/2$-rotation around the $Z$-axis followed by a $\pi/2$-rotation around the $Y$-axis.}
    \label{tab:logical_gates}
\end{table}

\section{Logical gates}
\label{sec:logicalgates}

Beyond the quantum memory capabilities that we have discussed so far, it is important to also consider operations within the logical subspace beyond single qubit Pauli operators. In general, it is hard to construct logical operators in a scalable way. However, for small near-term experiments, it is often useful to perform logical gates regardless to gain insight into potential practical constraints~\cite{marques2022logical, lacroix2025colorcode, egan2021faulttolerant, postler2022faulttolerantgates, ye2023logicalmagic, gupta2024magicstate}. For quantum LDPC codes, as constructed here, where the logical qubits are densely encoded, it is apriori not obvious how construct logical gates within the logical subspace. To demonstrate that logical gates are, in principle, possible, we follow the procedure from Ref.~\cite{kuehnke2025hardwaretailored} to numerically synthesize logical gates. This approach reduces the problem of synthesizing a logical Clifford operation to an integer quadratically constrained program. In Table~\ref{tab:logical_gates}, we show examples of logical Clifford operations on the logical qubits as defined in Eqs.~\eqref{eq:logical_z} and~\eqref{eq:logical_x}. More examples of logical gates can be found in Appendix~\ref{app:gates} including logical two-qubit gates between the two encoded qubits. These gate sequences for the logical gates were found with the \emph{htlogicalgates} package~\cite{kuehnke2025hardwaretailored} and are, at this point, proof-of-principles that logic can be performed in practice within the code-space. Large-scale architectures must however still be developed for example by relying on lattice surgery~\cite{horsman2012surfacecodelattice, litinski2019game} but adapted to quantum LDPC codes~\cite{yoder2025tourdegross}.

\begin{figure}[t]
    \centering
    \includegraphics[width=\linewidth]{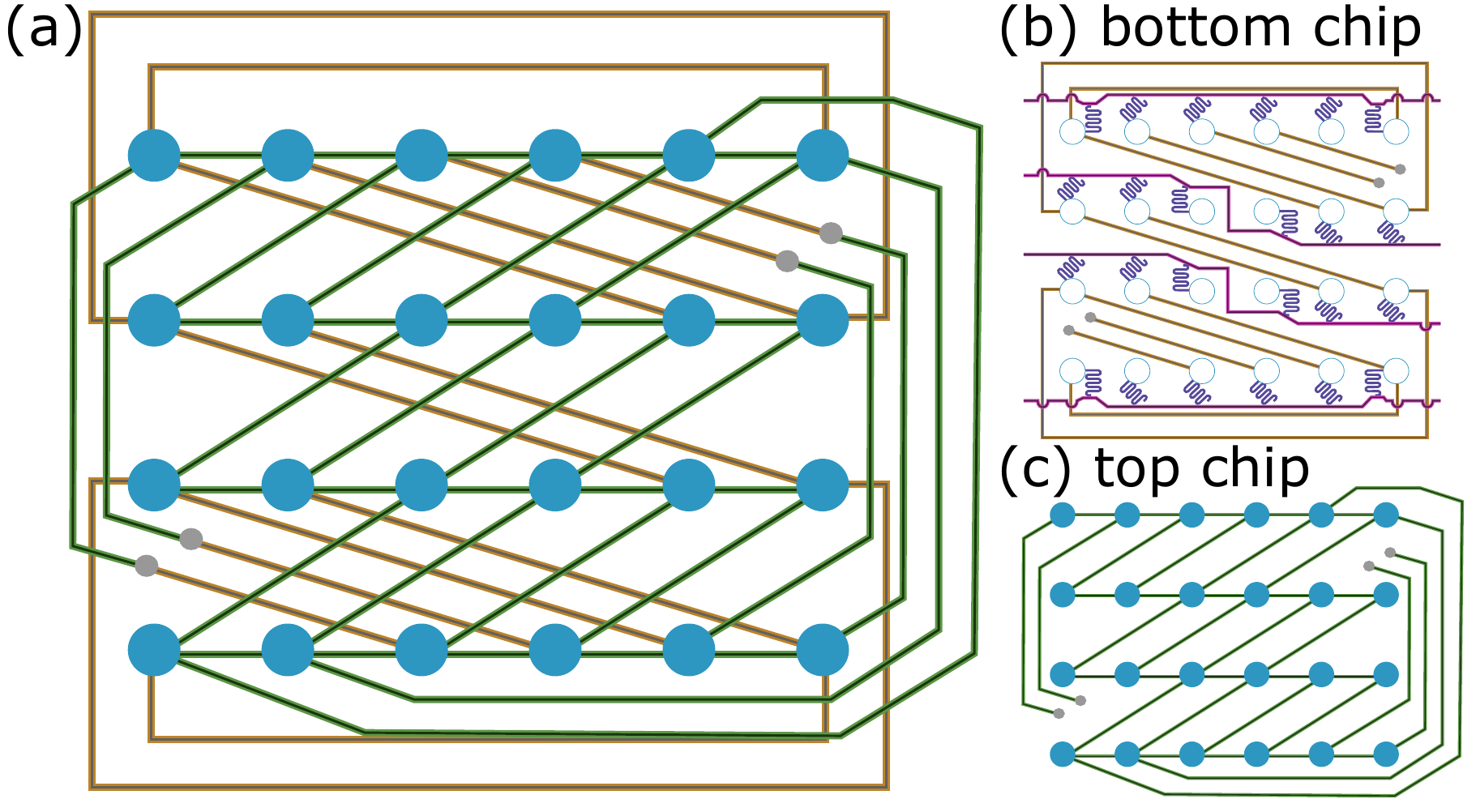}
    \caption{(a) Layout of the proposed quantum chip for superconducting circuits. The orange lines are couplers at the bottom chip in a flip-chip architecture while the green couplers sit on the top-chip. The small gray circles are bump-bonds connecting the bottom and the top chips. (b) Bottom chip only including the readout resonators for the qubits (purple) as well as the feedline for readout (pink). (c) Top chip only hosting the qubits and the green couplers.}
    \label{fig:sc_chip}
\end{figure}

\begin{figure}[t]
    \centering
    \includegraphics[width=\linewidth]{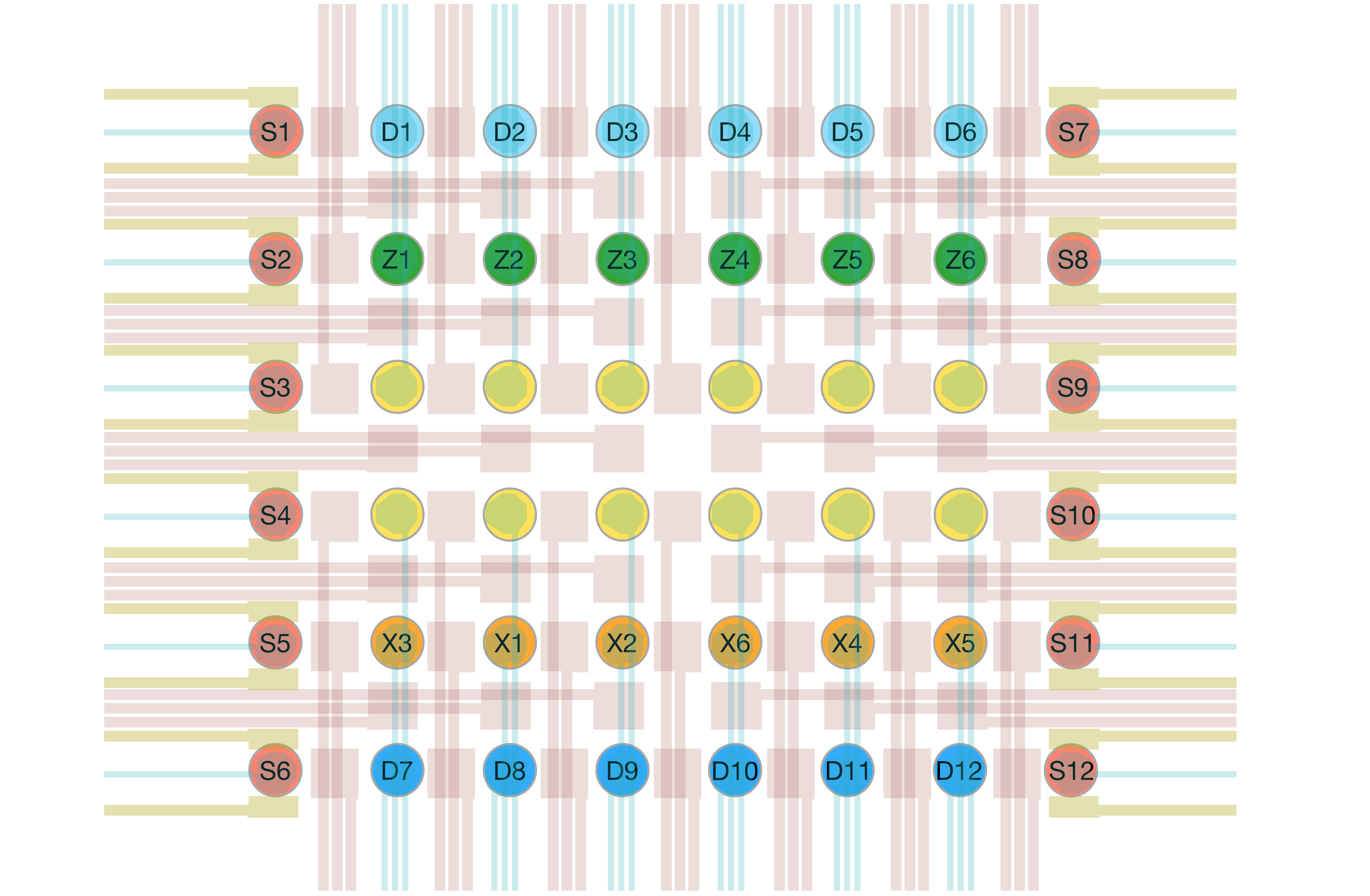}
    \caption{Illustration of a semiconductor device that may support the [[12,2,3]] code. We have indicated quantum dots in blue for the data-qubits, in green and orange for the auxillary qubits and in yellow for quantum dots used for shuttling of spin qubits. In red, we have additional quantum dots used as sensors for readout. Each quantum dot has a plunger gate in cyan and the quantum dots are seperated by barrier gates in red. Additionally, we have indicated the ohmic contacts in yellow for the sensor dots.}
    \label{fig:qd_chip}
\end{figure}

\begin{figure*}[t]
    \centering
    \includegraphics[width=\linewidth]{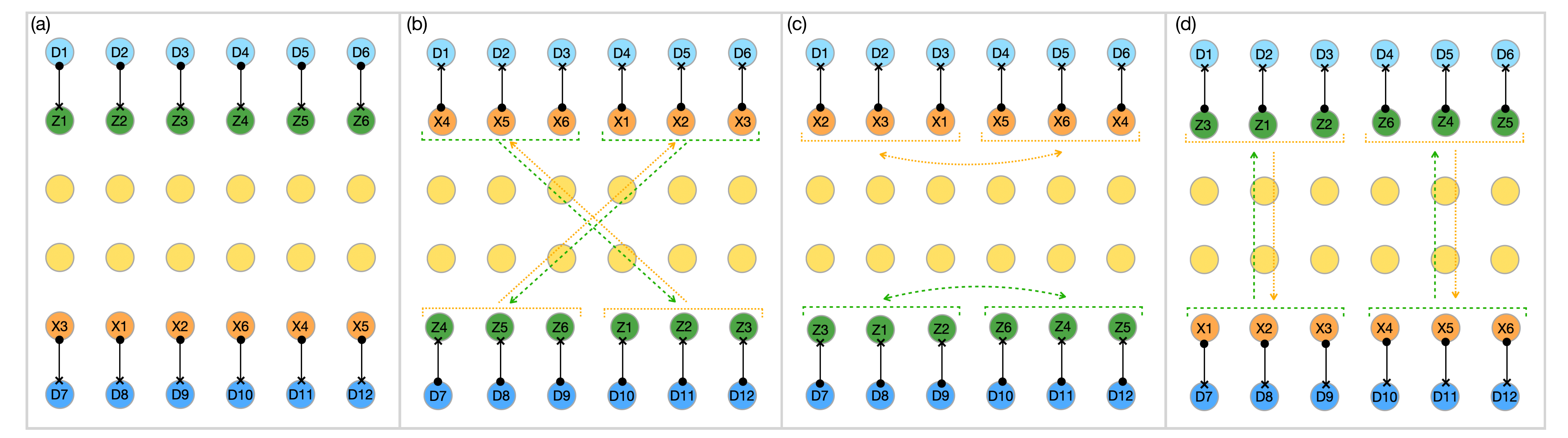}
    \caption{Key steps of the shuttling implementation of the [[12,2,3]] code: (a) The initial configuration has the data qubits in the outer layers directly connected to Z and X stabilizers respectively. The stabilizers can be moved in the group of three through the yellow shuttling dots. It is possible to always move two groups of stabilizers simultaneously. (b) Diagonal exchange of the two halves of the code in two consecutive steps followed by CNOTs on all qubits. (c) Horizontal exchange. (d) Vertical exchange.}
    \label{fig:qd_shuttling}
\end{figure*}

\section{Experimental implementations}
\label{sec:implementation}

Quantum error correction codes are only truly valuable if they can be experimentally realized. To experimentally realize any quantum error correction code that operates below the error correction threshold, each component must be optimized. Typically, we characterize the performance of quantum device through individual gate fidelities and two-qubit gate fidelities as well as readout and initialization fidelities. These low-level benchmarks are often done through techniques such as randomized benchmarking or similar benchmarks~\cite{knill2008randomized, magesan2011scalable, magesan2012interleavedrb, proctor2022mirrorbench, boixo2018characterizing, silva2024handsonrb}. However, at the end of the day, we need holistic benchmarks for the full device~\cite{proctor2024benchmarking}. While the ultimate benchmark would be running real large-scale algorithms, the most relevant intermediate benchmark is the performance of quantum error correction codes. Thus, when testing new hardware components such as long-range couplers or shuttling sequences needed for any quantum LDPC code, the most relevant benchmark is the performance of (small) quantum LDPC codes that utilize these new hardware components. As such, we will in the section present two proposals for near-term experiments that implement the small quantum LDPC codes presented above, specifically the $[[12,2,3]]$-code. Our proposals here focus on solid-state implementations but similar ideas can be tested with, for example, trapped ions~\cite{ryananderson2021realtime} or neutral atoms~\cite{bluvstein2023logical}.

\subsection{Superconducting qubits based on flip chip}

Superconducting qubits has been widely used as a platform to benchmark quantum error correction scheme~\cite{reed2012threequbitQEC, riste2015bitflip, kelly2015statepreservation, corcoles2015squarelattice, andersen2020repeated, acharya2022suppressing, google2025_belowthreshold, krinner2022repeated, marques2022logical}. Superconducting qubits are defined by electric circuits implemented with superconducting on low-loss substrates. Most superconducting qubits are of the transmon type~\cite{koch2007transmon} although there has been encouraging new progress with fluxonium qubits~\cite{manucharyan2009fluxonium, nguyen2019highcoherencefluxonium, ding2023fluxoniumftf, singh2025fastfluxoniumtransmon}. Interactions between superconducting qubits can be engineered by placing the qubits physically close to enable a capacitive or inductive interaction. Longer range coupling can be obtained through superconducting resonators in the context of circuit quantum electrodynamics~\cite{blais2021circuitqed}. However, to implement quantum LDPC codes, the long-range couplers are required to cross each other. Some degree of flexibility in the layout of a planar chip can be achieved with cross-over airbridges~\cite{versluis2017scalable}, however, care must be taken to not introduce additional crosstalk. More recently, there has been great progress in using additional wiring in a flip chip architecture including using two-qubit couplers across multiple chip-layers to perform entangling gates~\cite{kosen2022building, norris2025_multimodule, putterman2025hardware, conner2021flipchip, gold2021entanglementmodular}. 

Here, we propose to implement the $[[12,2,3]]$ code using a two-layer flip chip architecture, see Fig.~\ref{fig:sc_chip}. In this design, the couplers are distributed across two chip layers to avoid the need for coupler-crossings in a single plane. Specifically, as shown in Fig.~\ref{fig:sc_chip}(b,c), the orange lines indicate couplers fabricated on the bottom chip, while the green couplers are patterned on the top chip alongside the qubits themselves. The small gray circles represent superconducting bump bonds that electrically and mechanically connect the two chips, allowing signals to be routed vertically between layers.

Fig.~\ref{fig:sc_chip}(b) shows the bottom chip in isolation, which contains the readout resonators (purple) for each qubit and a shared feedline (pink) that enables frequency-multiplexed readout~\cite{heinsoo2018multiplexedreadout}. This configuration facilitates simultaneous measurement of multiple qubits without requiring individual readout lines for each, an important consideration for scaling to larger codes. Although the readout feedline does cross over the couplers in the layout, this is generally less problematic than having couplers cross each other, since the readout occurs at different frequencies than the two-qubit interactions.

The long-range couplers themselves could be realized as $\lambda/2$ resonators, where both the fundamental and second harmonic modes are exploited to mediate entangling interactions over larger distances. Alternatively, more innovative coupler designs, such as waveguides or multi-mode devices, could be integrated to further enhance flexibility and reduce crosstalk.

\subsection{Spin qubits based on electron shuttling}

The use of electron spins in quantum dots as qubits was one of the early ideas for building quantum computers~\cite{loss1998quantumdots, burkard2023semiconductor}. In particular, semiconductor quantum devices are especially promising due to their scalable fabrication which is compatible with semiconductor foundries~\cite{zwerver2022qubitmanufacturing, weinstein2023universallogic, neyens2024probing300mm, steinacker2024foundryspinqubit, george2025_12spin300mm}. Moreover, recent developments have led to high-fidelity readout and gates~\cite{blumoff2022fasttripleqd, takeda2024rapidssparity, noiri2022fastuniversal, xue2022quantumlogicspin, madzik2022precisiontomography, mills2022twoqubitsilicon}. However, scaling has in practice been challenging due limited flexibility in the readout design as well as the lack of long-range interactions. To overcome these challenges, recent effort has been put into enabling coherent shuttling of electron (or hole) spin qubits across the device~\cite{fujita2017coherentshuttle, mills2019shuttlingcharge, yoneda2021coherentspintransport, zwerver2023shuttlespin, vanriggelen-doelman2024germanshuttle, desmet2025highfidelity, katiraeefar2025evolutionary} which has also been proposed as a tool for surface codes~\cite{siegel2025snakes} as well as for quantum LDPC codes~\cite{micciche2025optimizing}. 

Here we propose a chip-layout for a spin qubit devices that relies on gate-defined quantum dots and uses shuttling to realize the non-local couplings needed for the quantum LDPC code. Specifically for the $[[12,2,3]]$ code, which requires 24 qubits, we propose a device with 36 quantum dots and an additional 12 sensor dots, see Fig.~\ref{fig:qd_chip}. The sensors are all placed at the edge of the device to enable good ohmic contacts. We organize the qubits into different rows of the device. We reserve the top row and the bottom row for the data qubits. Next we have two rows for the auxiliary qubits while in the center we reserve two rows to use during the shuttling operations. As also shown in Fig.~\ref{fig:qd_chip}, each quantum dot features a plunger gate and between each quantum dot, we place a barrier gate. Due to the partially overlapping gates, a careful tuneup of virtual gates must be performed before operating the device~\cite{vanDiepen2018automated, borsoi2024sharedcontrol, Rao_2025}. 

To operate the $[[12,2,3]]$ code, we must find a shuttling scheme that will enable the gate sequence in Fig.~\ref{fig:code_overview}(b). As illustrated in Fig.~\ref{fig:qd_shuttling}, we propose to carry out these operations by simultaneously shuttling three $X$-type and three $Z$-type auxilliary qubits while exchanging their order. For example, in Fig.~\ref{fig:qd_shuttling}(a) the qubits are in their initial configuration that allows for the first layer of CNOT gates. Next, in Fig.~\ref{fig:qd_shuttling}(b), we shuttle simultaneously the qubits $Z1$, $Z2$ and $Z3$ together with $X4$, $X5$ and $X6$. Using their buffer zone in the center two rows of the device, we can avoid any trajectory overlaps between individual qubits. This exchange of qubits can be followed by a similar exhange of the other two triplets of auxiliary qubits. Similarly, in Fig.~\ref{fig:qd_shuttling}(c), we use the buffer zone to re-order the auxiliary qubits along each their rows. Finally, in Fig.~\ref{fig:qd_shuttling}(d), the qubits are again shuttled in two pairs of triplets. After these four shuttling steps, the auxiliary qubits must be read out. The readout can be achieved by shuttling the qubits into the left-most and the right-most columns. During readout, we envision the data qubits to be stored in the two central columns of the device. Similarly for the final readout, we must fan out the data qubits to the left-most and right-most columns while storing the auxiliary qubits in the center. We further detail the shuttling sequence in a stop-motion-video, see Ref.~\cite{video}.

\section{Conclusion and Outlook}

In this work, we have presented a detailed review of small quantum LDPC codes as promising benchmarks and stepping stones toward scalable fault-tolerant architectures. Unlike local codes such as surface codes or 2D color codes, quantum LDPC codes offer the possibility of achieving lower overhead for a fixed performance, provided that the underlying hardware can deliver sufficiently low error rates and implement the required connectivity.

Following Ref.~\cite{voss2024multivariate}, we have introduced a family of bicycle codes, including a $[[12,2,3]]$ code and larger examples up to distance 7. These codes demonstrate favorable overhead compared to the rotated surface code while maintaining similar performance when decoded with a simple minimal-weight perfect-matching decoder. Our analysis reveals that the scaling of the code rate with distance improves over planar local codes. While the rotated surface code exhibits the expected $R \propto d^{-2}$ scaling, our constructions achieve a smaller exponent of approximately $1.7$, illustrating how non-local connectivity can help alleviate the footprint constraints that limit strictly planar and local layouts. At the same time, we emphasize that these benefits come with the practical challenges of engineering long-range couplers or qubit shuttling schemes, and of constructing appropriate code parameters for each target distance. 

To bridge the gap between theoretical constructions and experimental realization, we have outlined two concrete proposals for near-term implementations of the $[[12,2,3]]$ code. First, we described a flip-chip superconducting qubit architecture that distributes couplers across multiple chip layers to avoid planar layout constraints. Secondly, we proposed a spin qubit platform relying on coherent electron (or hole) shuttling to dynamically reconfigure non-local interactions within a compact footprint. Both approaches illustrate pathways to overcoming the scaling barriers of planar codes while retaining compatibility with existing fabrication processes.

More broadly, this work underscores that experimental demonstrations of small quantum LDPC codes are now within reach. Such experiments will not only provide valuable holistic benchmarks for quantum processors but also establish the groundwork for larger codes that exploit non-local connectivity to achieve higher rates and lower overhead. 

\section*{Code and data availability}
All the code and data to reproduce our results can be found at Ref.~\cite{GitHub}. The distance finding algorithm has been adapted from Ref.~\cite{shaw2025morphing}.

\section*{Acknowledgments}
We acknowledge useful input from M. H. Shaw and B. M. Terhal. The authors acknowledge the use of computational resources of the DelftBlue supercomputer, provided by Delft High Performance Computing Centre (https://www.tudelft.nl/dhpc).

\section*{Funding Declaration}
This research was supported by the European Union's Horizon Europe programme under the Grant Agreement 101069515 (IGNITE). C.K.A. acknowledges support from the Dutch Research Council NWO. 

\section*{Competing interests}
The authors declare no competing interests.

\section*{Author Contributions}
Both authors contributed to all aspects of conceptualizing, executing and writing of this work.

\appendix

\section{More Logical Gates}
\label{app:gates}

In the main text, we presented two examples of logical gates on the encoded qubits. In the Tables~\ref{tab:logical_cnot}, \ref{tab:logical_h2} and~\ref{tab:logical_s1}, we show examples of additional logical gates including an entangling gate between the two logical qubits.

\begin{table}[h]
    \centering
    \begin{tabular}{ll}
    \textbf{Logical gate} & \textbf{Physical gate sequence} \\
    \hline
    CNOT$_{L1,L2}$ 
      & (i) $X_2$ \\
      & (ii) $H_2$ \\
      & (iii) $Z_4$ \\
      & (iv) $X_4$ \\
      & (v) $C_4$ \\
      & (vi) $Z_5$ \\
      & (vii) $C_5$ \\
      & (viii) $Z_6$ \\
      & (ix) $C_6$ \\
      & (x) $X_{10}$ \\
      & (xi) $C_{10}$ \\
      & (xii) $H_1 \; H_3$ \\
      & (xiii) $C_{11}\; C_{12}$ \\
      & (xiv) $CZ_{1,2}$ \\
      & (xv) $CZ_{1,8}$ \\
      & (xvi) $CZ_{1,9}$ \\
      & (xvii) $CZ_{2,6}$ \\
      & (xviii) $CZ_{2,9}$ \\
      & (xix) $CZ_{2,12}$ \\
      & (xx) $CZ_{3,4}$ \\
      & (xxi) $CZ_{3,7}$ \\
      & (xxii) $CZ_{3,8}$ \\
      & (xxiii) $CZ_{3,10}$ \\
      & (xxiv) $CZ_{4,5}$ \\
      & (xxv) $CZ_{4,11}$ \\
      & (xxvi) $CZ_{5,9}$ \\
      & (xxvii) $CZ_{5,10}$ \\
      & (xxviii) $CZ_{6,7}$ \\
      & (xxix) $CZ_{7,8}$ \\
      & (xxx) $CZ_{7,12}$ \\
      & (xxxi) $CZ_{9,11}$ \\
      & (xxxii) $CZ_{10,11}$ \\
      & (xxxiii) $H_1 \; H_2 \; H_3$ \\
      & (xxxiv) $C'_4 \; C'_5 \; C'_6 \; C'_{10} \; C'_{11} \; C'_{12}$ \\
    \end{tabular}
    \caption{Logical $CNOT$ gates between the two logical qubits in the $[[12,2,3]]$ code. The gate $C'_i$ is a $\pi/2$ rotation along y-axis followed by a $\pi/2$ rotation around the x-axis.}
    \label{tab:logical_cnot}
\end{table}

\begin{table}[t]
    \centering
\begin{tabular}{ll}
\textbf{Logical gate} & \textbf{Physical gate sequence} \\
\hline
$H_{L2}$ 
  & (i) $X_1 \; X_2$ \\
  & (ii) $H_2$ \\
  & (iii) $X_4$ \\
  & (iv) $C_4$ \\
  & (v) $Z_7$ \\
  & (vi) $X_7$ \\
  & (vii) $X_{10}$ \\
  & (viii) $H_5$ \\
  & (ix) $C_9 \; C_{11}$ \\
  & (x) $H_{12}$ \\
  & (xi) $CZ_{2,7}$ \\
  & (xii) $CZ_{2,12}$ \\
  & (xiii) $CZ_{4,5}$ \\
  & (xiv) $CZ_{4,8}$ \\
  & (xv) $CZ_{4,9}$ \\
  & (xvi) $CZ_{5,6}$ \\
  & (xvii) $CZ_{5,8}$ \\
  & (xviii) $CZ_{5,9}$ \\
  & (xix) $CZ_{5,11}$ \\
  & (xx) $CZ_{6,8}$ \\
  & (xxi) $CZ_{6,9}$ \\
  & (xxii) $CZ_{7,12}$ \\
  & (xxiii) $CZ_{8,9}$ \\
  & (xxiv) $CZ_{8,11}$ \\
  & (xxv) $CZ_{9,11}$ \\
  & (xxvi) $C_2$ \\
  & (xxvii) $C'_4 \; C_5$ \\
  & (xxviii) $S_7 \; S_8$ \\
  & (xxix) $\sqrt{X}_9^\dagger$ \\
  & (xxx) $C'_{11}$ \\
  & (xxxi) $C_{12}$ \\
\end{tabular}
    \caption{Logical $H$ gate on logical qubit 2.}
    \label{tab:logical_h2}
\end{table}

\begin{table}[t]
    \centering
    \begin{tabular}{ll}
    \textbf{Logical gate} & \textbf{Physical gate sequence} \\
    \hline
    $S_{L1}$ & (i) $Z_2\; S_3 \; Z_4 \; S_5 \; S_7$ \\
     & (ii) $S_1 \; S_2 \; S_4 \; S_6 \; S_7$ \\
     & (iii) $CZ_{1,6}$ \\
     & (iv) $CZ_{8,9}$ \\
     & (vi) $CZ_{2,4}$ \\
     & (vii) $CZ_{3,5}$ \\
     & (viii) $CZ_{7,9}$ \\
     & (ix) $CZ_{7,8}$ \\
    \end{tabular}
    \caption{Logical $S$ gate on logical qubit 1.}
    \label{tab:logical_s1}
\end{table}

\bibliography{bibtex}

\end{document}